\documentclass[preprint2]{aastex}

\shorttitle{Multiple  supra--arcade downflows in turbulent solar flares}
\shortauthors{C\'ecere et al.}
\usepackage{amssymb}
\usepackage{natbib}
\usepackage{graphicx}
\usepackage{txfonts}
\usepackage{bm}
\usepackage[dvips,dvipsnames,usenames]{color}

\begin{document}
\title{3D MHD simulation of  flare supra--arcade downflows \\  in a turbulent current sheet medium}
 \author{M. C\'ecere\altaffilmark{1,2}, E. Zurbriggen\altaffilmark{1,2}, A. Costa\altaffilmark{1,2,3}, M. Schneiter\altaffilmark{1,2,3}}

 \altaffiltext{1}{Instituto de Investigaciones en Astronom\'\i a Te\'orica y Experimental
 IATE,  C\'ordoba, Argentina. }
  \altaffiltext{2}
 {Consejo Nacional de Investigaciones Cient\'\i ficas y T\'ecnicas (CONICET),
   Argentina.}
 \altaffiltext{3}{Facultad de Ciencias Exactas, F\'\i sicas y Naturales, Universidad Nacional de C\'ordoba (UNC),  C\'ordoba,  Argentina}

\begin{abstract}
Supra--arcade downflows (SADs) are sunward, generally dark, plasma density depletions  originated above posteruption flare arcades.
In this paper using 3D MHD simulations we investigate if the SAD cavities can be produced by a direct combination of the tearing mode and Kelvin--Helmholtz  instabilities leading to a turbulent current sheet (CS) medium or if the current sheet  is merely the background  where  SADs are produced triggered by an impulsive  deposition of energy.
We find that to give account of the observational   dark lane structures an addition of local energy,   provided by a  reconnection event, is required. We suggest that there may be  a closed relation between  characteristic SAD sizes and  CS widths that must be satisfied to obtain an observable SAD.

\end{abstract}

\section{Introduction}\label{Intro}

SAD features are known to be dark moving trails originated [$ 40-60$]Mm above eruption flare arcades with decelerating speeds  in the range $\sim [50 - 500] $km s$^{-1}$  \citep{2000SoPh..195..381M, 2009ApJ...697.1569M, 2011ApJ...730...98S}. They were first detected with the \textit{Yohkoh} Soft X--ray Telescope (SXT) \citep{1999ApJ...519L..93M}.  Since then, they have been extensively reported using other instruments such as \textit{TRACE} \citep{2003SoPh..217..267I, 2003SoPh..217..247I}, \textit{SOHO}/SUMER \citep{2003SoPh..217..247I} and \textit{SDO}/AIA \citep{2012ApJ...747L..40S}. There is  consensus in that due to  the lack of X--ray and extreme--ultraviolet (EUV) signatures in images and spectra, these SAD  structures are voided flows generated by reconnection processes in a CS above the flare arcade. 
 
 Several scenarios have been proposed to give account of the observations. After new AIA detections with high spatial resolution and temporal cadence, \citet{2012ApJ...747L..40S} re-interpreted SADs as density depletions left in the wake of thin shrinking loops. 
These authors indicated that either retracting loops  will be observed, when above the flare arcade is devoid of plasma (impulsive phase), or SADs, when a hot dense fan of plasma is present above the flare arcade (decay phase). They also proposed that deceleration is expected due to the buildup of downstream magnetic pressure and/or drag mechanisms.

Another scenario was proposed by \citet{2009EP&S...61..573L}, where the dynamic of retracting magnetic fields is triggered by a localized reconnection event that produces up and down flowing reconnected flux tubes, which are slowed down by underlying magnetic arcade loops. A drawback with this scenario is that the  observed SAD speeds are lower than expected for reconnection outflows in regions of typical Alfv\'en speeds of $1000$km s$^{-1}$. 

We are specially interested in the turbulent CS description given by \citet{2013ApJ...766...39M}. He  analyzed high--resolution observations in a sheet--like structure above a post--CME flare arcade where the  turbulent dynamic of a complex flow is described. He found that the plasma $\beta$ (the ratio of gas to magnetic pressure) is of the order of unity and described the flow variability in the hot plasma   ($T> 10$MK) as a product of  strong velocity shears and vortical motion where  small vortices moving towards the arcade were interpreted as probable SAD structures.

In \citet{2009MNRAS.400L..85C}, \citet{2010MNRAS.407L..89S}, \citet{2011A&A...527L...5M} and \citet{2012ApJ...759...79C} we reproduced the dynamics of multiple decelerating downflows through the assumption  that  dark tracks are confined  voided cavities --of high  $\beta$  and temperature values-- collimated in the direction of the ambient magnetic field and, generated by the bouncing and interfering of shocks and expansion waves upstream of the initial localized deposition of energy provided by reconnection events. In this scenario, the different observational SAD sizes  could be interpreted   either  as  the consequence of reconnection events that are triggered in a  homogeneous background,  or as  the consequence of reconnection events produced in a previously distorted media by the passage of earlier  SADs. We found that the observed wavy character  \citep{2005A&A...430L..65V} can be interpreted as  an indication of interaction between SADs. This interaction is significant when the bursts that trigger the phenomenon act on the wakes left by previous SADs.

 Recent observational data and modeling have challenged the scenario described by \citet{2012ApJ...759...79C}. \citet{2014ApJ...786...95H} measured the plasma temperature of the SAD regions and surrounding plasma sheet using AIA and XRT data. They calculated differential emission measures for several flares and their corresponding SADs and found that there is little convincing evidence to sustain the high temperatures in the SADs predicted by \citet{2011A&A...527L...5M} and \citet{2012ApJ...759...79C}. They also found that  SADs are always hotter than the background, but in many cases  cooler than the surrounding fan plasma. 
 
 Related with thermal conduction considerations, one of the major challenges is to understand how it is possible that SADs can last  in hot CS. In fact, structures with typical SAD sizes of decades of Mm, typical coronal number densities of $n \approx 10^{9}$cm$^{-3}$ and  temperatures as high as $T\approx 10$MK  \citep{2013ApJ...766...39M} will vanish in times (of a few seconds) that are at least two orders of magnitude lower than the observed values. However, we  show that  considering  a typical coronal background ($T\sim  1$MK) or/and the high density fan region ($n\sim 2\times 10^{10}$cm$^{-3}$) as the   medium where the SAD dynamic develops, the thermal conduction effects are lower enough to allow the comparison with the observations. 
 
 In what follows motivated by the description provided by \citet{2013ApJ...766...39M} we   explore a new  scenario.  We  consider a quasi--2D turbulent CS as the medium where  SAD features can be  observed. We propose that SADs are voided cavities formed  by nonlinear waves. Waves which are triggered by  bursty reconnection events (blast wave  expansion mechanism, \citet{1988SoPh..117...97F, 2013SoPh..288..255K})  that occur during  a larger scale reconnection process. This is,   the quasi--2D turbulent CS evolution times are much larger than the SAD ones. We emulate these individual  reconnection events by pressure pulses. 

We perform
3D MHD simulations including magnetic resistivity and assuming that heating and  cooling terms compensate each other. The paper is structured as follows: In Sec. \ref{TCS} we consider    turbulence and  CS formation; in  Sec. \ref{TC} we justify   the assumption made regarding the cooling (conduction and radiation) and heating (reconnection) term; in Sec. \ref{Model} we present the model; in Sec. \ref{nsup} we state the numerical setup and initial conditions.  In Sec. \ref{RD} we discuss  the results obtained and in Sec. \ref{c} we summarize the conclusions. 

\section{Reconnection processes: turbulent current sheets and bursty reconnection} \label{TCS}

The notion of quasi--separatrix layers proved to be fertile since   observational and theoretical  reconnection studies showed that 3D reconnection requires  to go beyond the classical generalization of 2D null points and their correspondent separatrices (e.g. \citet{1995JGR...10023443P,1997A&A...325.1213S}). 

However, the study of 2D CS formation and evolution is still important even for 3D calculations. \citet{2004PhPl...11.4837O} simulated magnetic reconnection on a slab geometry and showed that the presence of a global guiding magnetic field makes the 3D evolution much similar to those of purely 2D (see also \citet{2010PhPl...17a0702F}). The formation of smaller scale structures associated with a direct energy transfer, allowing larger diffusion and faster reconnection rates, is a consequence of the nonlinear evolution that eventually ends in a turbulent regime of many spatial scales. The simulations with a guide field produce both, a  direct and an inverse energy cascade. The inverse energy transfer generates coalescence of magnetic islands which are typical 2D structures. In the direct cascade the wavelengths decrease with increasing distance from the CS.  Coalescence is mostly suppressed when  faster 3D phenomena are significative. Thus, the fact that coalescence is present is an indication that a 2D picture is a good description of a phenomenon.  

Typical observational evidences of these  quasi--2D CS configurations were described by \citet{2013ApJ...771L..14G} where the coalescent plasmoids,  or those  long--lasting slab configurations usually seen  above arcades  can be observed (e.g. \citet{2000SoPh..195..381M, 2009ApJ...697.1569M, 2011ApJ...730...98S}).

There is a broad range of  determinations concerning the thickness of 2D CS. \citet{2013ApJ...771L..14G}, based on observational data  and using a 2D simulation  estimated an upper limit of the CS width of $3 $Mm.  However, associated with non-thermal line widths, \citet{2008ApJ...686.1372C} reported  CS thickness ranging within $\sim [28-56]$Mm and  \citet{2008ApJ...689..572B} attributed the large temperatures and observed CS thickness ($\sim [10-100]$Mm)  to turbulence. Discrepancies between widths could be due to, either the impossibility to observationally distinguish between the CS and a sheath of hot plasma surrounding the CS \citep{2009ApJ...701..348S}, to non-thermal bulk flows and/or turbulence \citep{2008ApJ...686.1372C,2008ApJ...689..572B} or to  different regimes of 2D CSs behavior \citep{1978A&A....64..219H} where an adiabatic description would be accurate.

Deviating from these 2D descriptions, flares that release impulsive energy in absence of a sustained gradual phase  were extensively studied  (e.g. \citet{1982QB539.M23P74...}).  The existence of explosive events in reconnection processes imply that they occur on a faster timescale than  large--scale ones. It was suggested that this could be either due to the presence of stressed magnetic flux tubes that become unstable and produce individual bursty reconnection in the frame of the  longer term process or because reconnection itself is inherently impulsive and bursty \citep{1986SoPh..104....1P,2000mare.book.....P}. \citet{1988SoPh..117...97F}  pointed out the importance of blast waves produced by pressure--driven expansions and shocks due to impulsively driven reconnection, and recently \citet{2013SoPh..288..255K} observed a limb flare and proposed a flare blast wave scenario to describe it. 

Thus, due to the above discussion   we will consider a turbulent quasi-2D CS of different  thicknesses as the medium 
where SADs may be observed.

 \section{The effects of thermal conduction} \label{TC}

\citet{2009ApJ...701..348S} studied a CS  model to analyze reconnection outflow jets considering  thermal conduction.
Based on a model by   \citet{1987INTSA..34..136S} they assumed that a 2D CS is generated by a Petschek-type reconnection model and found that, for large  heat conduction values, the internal CS temperatures  are  almost uniform and equal to the background values  (see Figure 7 in \citet{2009ApJ...701..348S}).  Considering that surrounding a CS an expanded  thermal halo is formed, they assumed  an almost steady configuration where, in accordance with observations, density is uniform and  internally enhanced over the background coronal densities \citep{2006SSRv..123..127S}. However, this  CS scenario of almost uniform density and low temperature,  dominated by heat conduction, does not seem to be  consistent with a typical  turbulent inhomogeneous hot media where  SADs are observed. It seems that only when conduction is not dominant with respect to the reconnection process (they did not consider cooling by  radiation), higher  internal CS temperatures and nonuniform density distributions would be obtained.

In fact, \citet{2008ApJ...686.1372C} studied the CS associated with the $2003$ November $4^{th}$ CME and found  large non-thermal [Fe XVIII] line broadening. The corresponding speeds were as high as $380$km s$^{-1}$ at an early stage, and later -in a fairly constant phase-  they ranged between $[50-200]$km s$^{-1}$. They concluded that these non-thermal effects are explained by the presence of turbulence and bulk flows.

Although conduction and radiation are both  cooling processes, while conduction proceeds to distribute heat --being highly efficient in hot CSs--  the radiation acts as a sink  function that  takes heat away locally. Thus, the heat conduction tends to expand the system distributing the energy, and the radiation tends to narrow it by reducing the gas pressure with respect to the surrounding media. Theoretical and observational CS studies (e.g. \citet{2008ApJ...689..572B, 2008ApJ...686.1372C}) support the existence of CSs with quasi stable thickness values, thus, reconnection processes should provide the energy to stabilize the CS width for times comparable with the observations. 
Thus, 
a certain set of CS parameters could give account of an   energy balanced open system where diffusion is limited allowing the observation of SADs for times comparable with the observations. 

 There are three main physical processes for energy balance in the CS: thermal conduction, radiation and reconnection. The timescales of conductive and radiative cooling   are:
\begin{equation}
t_{cond} \equiv \frac{3 n k_B L^2}{\kappa_0 T^{5/2}}, \quad
t_{rad}   \equiv \frac{3  k_B T}{n E_r}, \label{1}
\end{equation}
where $n$ is the number density, $k_B$ is the Boltzmann constant, $L$ is a characteristic length,  $\kappa_0 \approx 10^{-6}$erg K$^{-7/2}$ cm$^{-1}$ s$^{-1}$ is the heat  conduction coefficient along the magnetic field, $T$ is the temperature and $E_r$ is the radiative loss function ($E_r\simeq 4\times 10^{-23}$erg cm$^{3}$ s$^{-1}$ for $T\backsimeq 10$MK) \citep{2005psci.book.....A}.

To evaluate the timescale of reconnection heating, $t_{rec}$, we follow \citet{1999ApJ...517..700L} who proposed a low dissipation model where a turbulent quasi 2D CS is formed due to the presence of a guiding magnetic field that reduces the transverse scale for reconnection flows. Turbulent motion create small regions of intense field gradients and reconnection occurs in small layers spread throughout the larger direction. Associated with their model the reconnection  timescale is:
\begin{equation}
t_{rec}  \equiv \frac{L}{v_A M_A^2} \label{2}
\end{equation}
where $v_A$ is the Alfv\'en speed and $M_A$ is the Mach Alfv\'en number. 

Hence, our aim is now to determine realistic CS physical parameters which lead  to timescales compatible with an almost non-diffusive description, i.e.,  where SAD features can develop for times comparable with the observations. 

\section{The Model: SADs as blast waves explosive events during a long duration reconnection process in a turbulent media} \label{Model}

\begin{figure}[htb!]
\centering
 
  \includegraphics[width=8.cm]{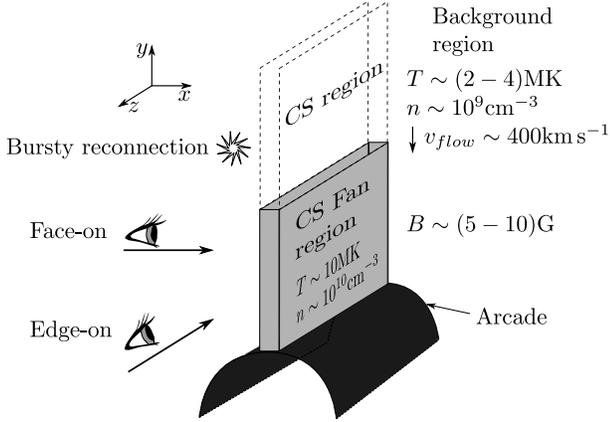}
  \caption{Simplified scheme of  the 2D CS, formed by stochastic reconnection \citep{1999ApJ...517..700L},  immersed in a coronal  background media. A blast-wave   reconnection process \citep{2013SoPh..288..255K} leading to the formation of a SAD, is also indicated. }\label{fig:1} 
 \end{figure}

We assume that 
 SADs are voided and expanded cavities resulting from  bursty   reconnections \citep{1988SoPh..117...97F, 2013SoPh..288..255K} that occur during  a larger scale 2D turbulent CS reconnection process. The triggering blast wave mechanism was proposed in \citet{2009MNRAS.400L..85C} (see e.g., the explanation given in \citet{2012ApJ...759...79C} and Figure~3 therein).    
  They are assumed as local  features independent of the overall turbulent  quasi--2D CS. Thus, they may eventually be triggered  outside the CS reconnection region. However, because of  the simplicity of the  setup used (where the background  is not modeled, only the CS environment was used to analyze the dynamic behavior)  we  can only simulate the  case where  SADs are triggered (already immersed) inside the CS, but we will  argue, in the conclusion section, about what could happen when they are triggered  outside the CS (see Figure~\ref{fig:1}).

The initial conditions for the chosen CS parameters are: an average temperature value of $T = 10$MK,  an initial sunwardly guiding magnetic field value
of $B=5.9$G  ($y$ axis), and an enhanced CS  number density value of $n=2 \times 10^{10}$cm$^{-3}$. 
As  heat conduction is strongly inhibit across the  magnetic field lines, and considering that the  guiding magnetic field is oriented in the $y$ direction  we calculate the heat conduction timescale considering a typical CS length of $L=140$Mm. 
With these values and using Eq.~(\ref{1})
we obtain:
$ t_{cond}\simeq t_{rad} \simeq 5100 $s. 

 With the above CS parameters, the Alfv\'en speed results $v_A=92$km s$^{-1}$ and,  considering an average turbulent speed of $v=50$km s$^{-1}$ ($M_A\sim 0.5$) \citep{2008ApJ...686.1372C,2013ApJ...766...39M}, using Eq.~(\ref{2}) we obtain $ t_{rec}\simeq 5247$s. 
 
If we consider that a typical SAD is triggered by a blast phase associated with an initial  adiabatic expanding shock, thermal conduction will not play a significant role in the exchange of heat with the surroundings at this early stage. Later, the thermal exchange with the SAD neighborhood will strongly depend on the field orientation.  The blast stage leads to a magnetic configuration where the magnetic field lines tend to displace and envelope the SAD, thus the tendency is to  thermally insulate it from the surroundings. 
Meanwhile, considering that  SADs  travel    $\sim (20-40)$Mm along the sunward $y$ direction (appearing  as  elongated feature of $\sim 40$Mm in the fan region) we respectively obtain:  $ t_{cond}\simeq (105-400)$s, which is consistent with  observational times.  See e.g., the movies provided by  \citet{2012ApJ...747L..40S}.

With these assumptions, taking into account the calculated timescales, we  consider that the cooling and heating terms compensate each other and that the conduction will not substantially alter the results for times comparable with the SAD observations. Different values of the chosen physical parameters would lead to a different scenario. In favour of this one we can argue that lower thermal conduction times associated with other realistic set of parameters would lead to a rapid extinction of the  observed features. We speculate that this  particular scenario could explain why SADs are not always observed associated with  supra--arcade CSs. A model considering cooling terms due to anisotropic thermal conduction, radiation, and reconnection heating is in progress.

  \section{Numerical code and initial conditions}  \label{nsup}

We numerically solve the MHD equations. In conservative form they read  (CGS units): 
\begin{eqnarray}
\frac{\partial \rho}{\partial t} + \nabla \cdot \left(\rho\mathbf{v}\right)  \!\!\!&=&\!\!\! 0, \label{3} \\ 
\frac{\partial \rho  \mathbf{v}}{\partial t} + \nabla \cdot \left(\rho\mathbf{v}\mathbf{v}-\mathbf{B}\mathbf{B}\right) + \nabla p_* \!\!\!&=&\!\!\! 0, \label{4} \\ 
\frac{\partial e}{\partial t} + \nabla\cdot \left[(e+p_*) \mathbf{v}-\mathbf{B}(\mathbf{v}.\mathbf{B})\right] \!\!\!&=&\!\!\!  \nonumber \\ 
\nabla \cdot \left[\mathbf{B} \times (\eta \nabla \times \mathbf{B}) \right],  \label{5} \\ 
\frac{\partial \mathbf{B}}{\partial t} + \nabla\cdot \left(\mathbf{v}\mathbf{B}-\mathbf{B}\mathbf{v}\right) \!\!\!&=&\!\!\! \eta \nabla^2\mathbf{B}, \label{6} \\ 
e = \frac{1}{2}\rho v^2 + E+\frac{B^2}{2},  &&  \label{7} \\ 
p_* = p+\frac{B^2}{2}, && \\
 p=(\gamma -1)E, && \label{8}
\end{eqnarray}
where $\rho$ is the mass density, $t$ is the time, $\mathbf{v}$ is the plasma flow velocity, $p$ is the thermal pressure, $\mathbf{B}$ is the magnetic field divided by $\sqrt{4\pi}$, $e$ the total energy, $E$ the internal energy, $\gamma=5/3$ the rate of heat coefficients and $\eta$ is the resistivity.

The software used in this work was in part developed by the ASC/Alliance Center for Astrophysical Thermonuclear Flashes at the University of Chicago \citep{2000ApJS..131..273F}.
    We perform  3D MHD simulations with the extensively validated FLASH4
  code using  the adaptive mesh refinement procedure (AMR) with the Powell's $8$-wave scheme  \citep{1999JCoPh.154..284P} to solve the  MHD equations.

   To generate a turbulent CS we start from a  monolithic CS configuration with Lundquist value   $S \gtrapprox 10^{10}$, greater than the critical value  ($S_{c} \sim 10^{4}$) at which the Sweet--Parker CS becomes unstable.
 In this way, a turbulent regime is generated  giving  account of a hierarchical configuration of  overdense plasma features connected by secondary CSs as  described in the literature  (see references in \citet{2000mare.book.....P}).  
  The initial regime is set up  with a uniform diffusivity of $\eta \approx 1 $m$^{2}$s$^{-1}$,
   which is larger than the Spitzer value for a typical 
  CS
  background temperature of $T= 10$MK  ($\approx 0.03$m$^2$s$^{-1}$, $\eta\sim 10^{9}T^{-3/2}$m$^{2}$s$^{-1}$ with  $[T]=$MK),  yet lower than the estimated anomalous diffusivity \citep{2008ApJ...689..572B}.
  We assume  the  CS parameters discussed above: $n=2\times 10^{10}  $cm$^{-3}$, $T= 10$MK and $B_0=5.9$G. A Cartesian grid with $4$ levels of refinement was employed leading to a maximum grid refinement of $(128,256,64)$. The  physical domain was set up to  $(50,100,25)$Mm (see right bottom corner of Figure~\ref{fig:2}), with  the $y$ coordinate pointing  
 sunwards  and the $x$ and $z$ coordinates perpendicular 
 to the initial magnetic field direction. The basic CS device is initialized  assuming a corona in pressure equilibrium with a   velocity perturbation  in the $x$ direction given by
  \begin{equation}
pert =v_{0} \sin(2\omega y)\times random,  
\ \ \ \omega=\frac{2\pi}{y_{max}-y_{min}},  \label{9}
\end{equation}
  where $v_{0}=2$km s$^{-1}$, $random$ is a uniform distribution  of  numbers 
   (varying between $(0-1)$) and $y_{max}-y_{min}$ is the domain size in the $y$ direction.
    In addition to this perturbation, we consider various CS models with initial perturbed shear velocities (at $t=0$s) in the $y$ and $z$ directions (see Table~\ref{tab:table1}).
  The  magnetic field configuration is:
\begin{equation}
B_{y} = \left\{
\begin{array}{l l}
 \ \   B_{0}  & \quad \mbox{if $ x < 0 $}\\
 -B_{0} & \quad \mbox{if $ x \ge 0$ }
\end{array} \right.  \label{10}
\end{equation}
where $B_{0}$ depends on the model. Periodic boundary conditions are assumed in the direction where a shear  will be imposed  (a $\mathbf{K}$  variable satisfies in direction $n$:   $\mathbf{K}(n_{ min})=\mathbf{K}(n_{max})$  because no boundary effects are expected in the development of turbulence), otherwise outflow conditions are assumed. 
\begin{table*}
\begin{centering}
\begin{tabular}{@{}cccccc}
Model & $B_{0} [G]$ &  $v_{x}[km \ s^{-1}]$ &  $v_{y}[km \ s^{-1}](x<0|x>0)$ & $v_{z}[km \ s^{-1}](x<0|x>0)$ & $(\Delta P/P)$ \\ \hline  
$M0$ & $5.9$  & $pert$ & $0 | 0$                     & $0 | 0$                  & $0$ \\ \hline
$M1$ & $5.9$  & $pert$ & $ (0 | -364) + pert$ & $0 | 0$                  & $0$ \\ \hline
$M2$ & $0.59$ & $pert$ & $0 | 0$                     & $(0 | -182)+pert$ & $0$ \\ \hline
$M3$ & $5.9$  & $pert$ & $0 | 0$                     & $(0 | -364)+pert$ & $0$ \\ \hline
$M4$ & $5.9$  & $pert$ & $(0 |-364)+pert$     & $0 | 0$                  & $4$ \\ \hline
 \hline
\end{tabular}
\caption{\label{tab:table1} Simulated models:  $B_{0}$ is the background magnetic field, $v_{i}$ is the initial velocities in the $i$ direction, $x=0$ is the position of the CS, $pert=v_{0} \sin(2\omega y)\times random $ is the perturbation (same functional $y$ dependence for all directions).}
\end{centering}
\end{table*}

\section{Results and Discussion}\label{RD}
\subsection*{A turbulent picture}
\begin{figure}[htb!]
\centering
   \includegraphics[width=6.cm]{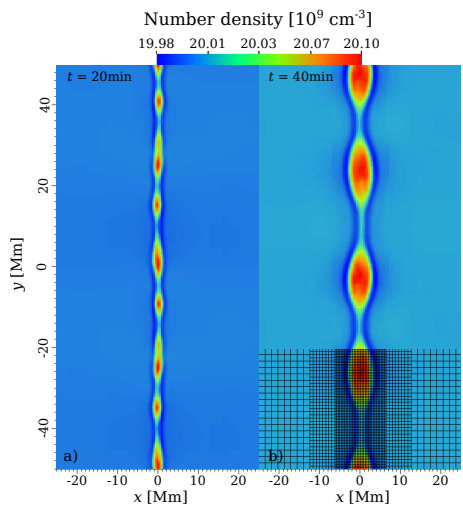}
 \caption{Simulation of $M0$, density slices of the plane $z=0$ at a) $t=20$min and b) $t=40$min, superimposed we show the refinement of the grid.}\label{fig:2} 
 \end{figure}

From the initial conditions of the coronal plasma parameters (model $M0$ see Table~\ref{tab:table1}) we obtain  a turbulent configuration.
Density slices of the $z=0$ plane (edge-on view) are shown in  Figure~\ref{fig:2}a--b for   times: $20$min and $40$min, respectively. As shown in the figure, and extensively described in the literature, the tearing mode  instability leads to a turbulent regime composed of  dynamic and coalescent plasmoids where the desired subdense structures  are only obtained as  secondary linear CSs (they connect neighbor plasmoids).
Thus, the usual tear--drop--shaped SAD features are not easy to obtain  from models of the $M0$ type in times comparable to the observations.  

 However, if  an instantaneous  shear in the flow speed to the sides of the CS is introduced at $t=0$s (models $M1$, $M2$ and $M3$ of Table~\ref{tab:table1}), the turbulent features change markedly allowing the appearance of subdense cavities.
 It is well known that Kelvin--Helmholtz perturbations  destabilize  CSs  \citep{2008ApJ...689..572B}. When a shear  initiates  Kelvin--Helmholtz perturbations and  combines with the tearing instability,  the overall dynamic is modified.  
 Accordingly, \citet{2013ApJ...766...39M} reported  strong coronal velocity shears (up to $\simeq [250-350]$km s$^{-1}$).
  Also,  as  large sunward flow values ($v_{flow}\simeq v_{A}\simeq [300-500]$km s$^{-1}$, $v_{A}$ is the external Alfv\'en speed\footnote{$ v_{A}$ is the background  Alfv\'en speed, where  density and temperature have typical coronal values, e.g. considering $n=  10^9$cm$^{-3}$ and $B_0=5.9$G, $v_{A}= 414$km s$^{-1}$.})  are expected   coming from a reconnection  site,  strong shears in the flow could arise  due to e.g. the  inhomogeneities  of the flaring medium  (see e.g. the  flow speed coming from the right side into the fan structure in 
 slices between $11:58:09$ and $12:02:09$  of movie $1b$ by \citet{2012ApJ...747L..40S}).
 
\begin{figure}[htb!]
\centering
  \includegraphics[width=6.cm]{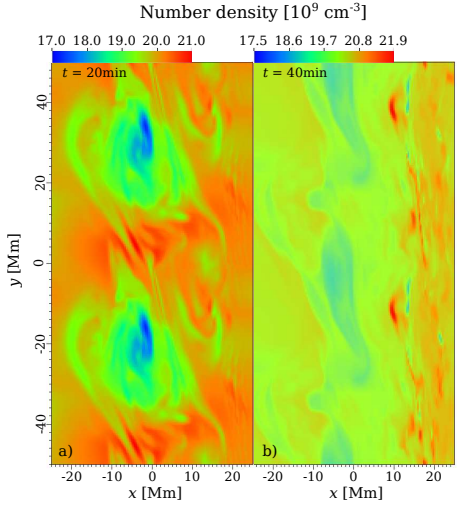}
 \caption{Simulation of $M1$, density slices of the plane $z=0$ at a) $t=20$min and b) $t=40$min.}\label{fig:3} 
 \end{figure}

 We thus perform several runs with initial strong shear velocities (see Table~\ref{tab:table1}) to gain insight into the turbulent features.  
 Figure~\ref{fig:3} shows density slices of the $z=0$ plane  for $M1$, and for the same  times as in Figure~\ref{fig:2}. The model configuration is the same as  in $M0$ with the  addition of a random speed with a shear in the $y$ direction. Comparing Figure~\ref{fig:3} and Figure~\ref{fig:2} we note that models with shear can generate subdense cavities, lasting for times of the order of  decades of minutes as  the main features of the turbulent regime.


\begin{figure}
\centering
 
  \includegraphics[width=6.cm]{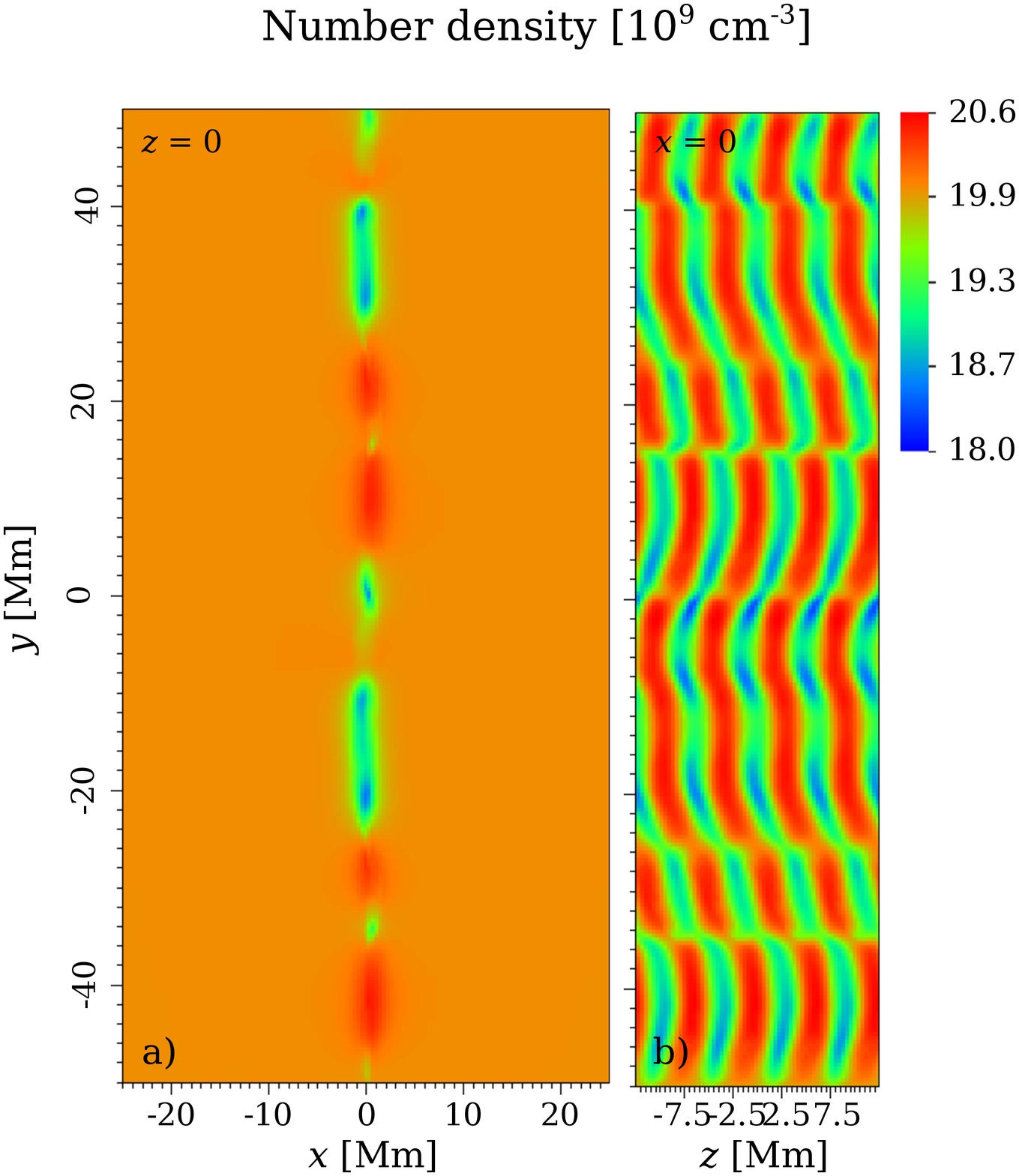}
    \includegraphics[width=6.cm]{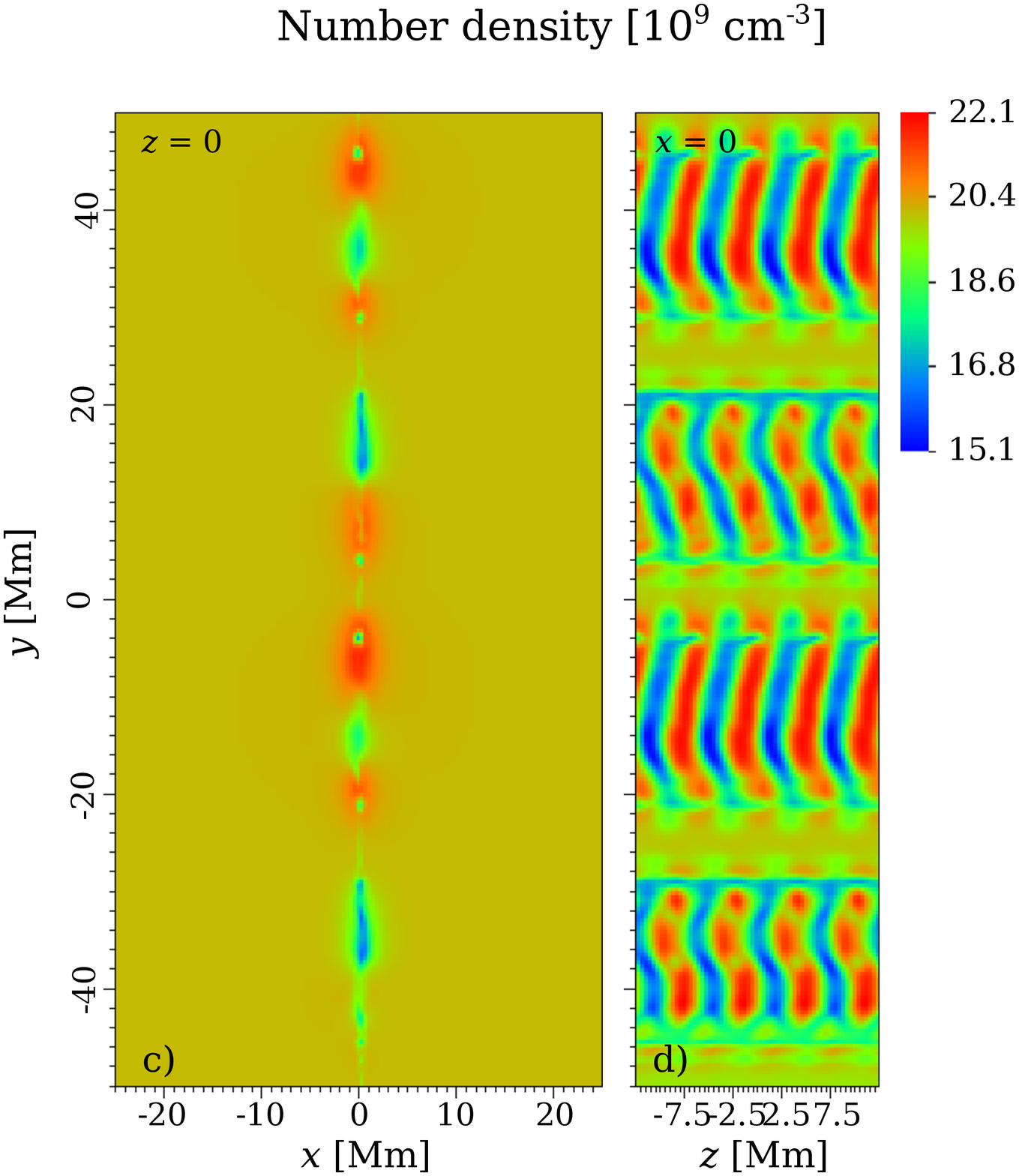}
 \caption{Simulation of model $M2$ at $8$min, density slice: a) Edge-on view ($z=0$) and b) Face-on view ($x=0$). Simulation of model $M3$ at $t=4.5$min, density slice: c) Edge-on and d) Face-on.}\label{fig:4} 
 \end{figure}
 
 We also performed  runs  with a shear in the $z$ direction (models $M2$ and $M3$).
  As in  Figure~\ref{fig:3}, Figure~\ref{fig:4}a and Figure~\ref{fig:4}c show evolving subdense structures. 
The viewing orientation of Figure~\ref{fig:4}b and Figure~\ref{fig:4}d is, face-on, perpendicular to  the CS plane. These $x=0$ plane descriptions  resemble observational inhomogeneities  as seen, for example,  in movies $1a$ and $1b$, or figure $1$ by \citet{2012ApJ...747L..40S}. Nevertheless, these  sunward subdense features are not stable during times comparable with the  observation of that inhomogeneities, they cannot sustain their shape for more than $1.2$min.

\subsection*{Later energy depositions}

 Once the turbulent CS is developed, we apply an instantaneous pressure pulse to emulate a blast 
   reconnection event \citep{1988SoPh..117...97F}
   occurring high in the corona where reconnection
   is prone to occur, triggered by a local change in the magnetic field
   line linkage and the magnetic topology, e.g., null points,
   separatrices or 3D quasi--separatrix layers, (see
   \citet{1996A&A...308..643D}). Such initially localized processes that lead to
   bursts of impulsive deposition of energy that evolves producing an  expansion of the plasma have been proposed in
   \citet{2013ApJ...776...54S, 2012ApJ...759...79C,
     2011A&A...527L...5M, 2009MNRAS.400L..85C}.
       Also, the reconnection flare blast wave scenario was observationally confirmed by \citet{2013SoPh..288..255K}, revealing
    the formation of initially expanding cavities that later  collapse inwards while they approach an  arcade, i.e., as they reach a supposedly denser medium. 
    
      \begin{figure}[htb!]
\centering
  \includegraphics[width=6.cm]{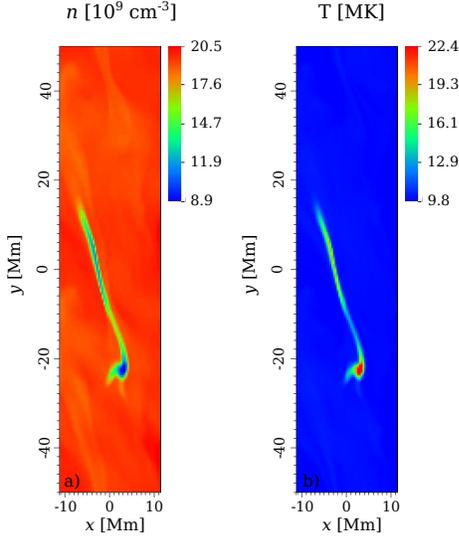}
  \caption{Slices of the $z=0$ plane for Model $M4$ with a pressure pulse diameter of $4$Mm, at $t=50.3$min: a) number density, $n$ and b) temperature, T. See also the movies attached to the electronic version of the paper. }\label{fig:5} 
 \end{figure}
 
 Figure~\ref{fig:5}a--b shows one of these events
    (model $M4$) resulting from an instantaneous  spherical pressure pulse that is four times its background value\footnote{This increase in the pressure can be produced e.g. by a burst reconnection of a stressed magnetic flux tube. To estimate the pressure pulse note that the temperature of a flaring loop can be as large as $40 $MK \citep{2005psci.book.....A}, and  assuming  that 
  the density of the  pulse is initially the same as the environment one, we obtain    $ \Delta P /P=\Delta T/T=4$, considering  that the fan CS temperature is of $\approx 10$MK. The pressure pulse value could  be larger if an increase  of the density is also allowed, e.g.,  the    number density of a flaring loop can be as high as $n\approx 10^{11}$cm$^{-3}$, much larger than the fan  value considered.}  at $t=46.1$min.       
    We assume a perturbation diameter of  $d=4$Mm located at $(0,35,0)$Mm   leaving the medium density unaltered and allowing the increase of the   internal temperature. The
    figure shows the $z=0$ slice for the number density and the
    temperature at $t=50.3$min. We obtain a  tear--drop
    SAD that travels  sunwards a distance of $\sim 60$Mm with a speed of $\sim 240$km s$^{-1}$,
    leaving a persistent voided region   along a distance of $\sim 38$Mm.
     The fan  number density  is at least twice that of the
     SAD, and the SAD temperature is $22.4$MK, whereas the number
     density of the turbulent  background    (Figure~\ref{fig:3}) is 
     only $\sim 1.23$ times the eddy subdense cavities of  temperature  $10$MK. 
  In accordance with the observations
  \citep{2013ApJ...766...39M} our simulations show that the turbulent background
  has  $\beta\geq 1$ values. 
 
 \section*{Dynamic behavior}
 
  \begin{figure}[htb!]
\centering
   \includegraphics[width=8.cm]{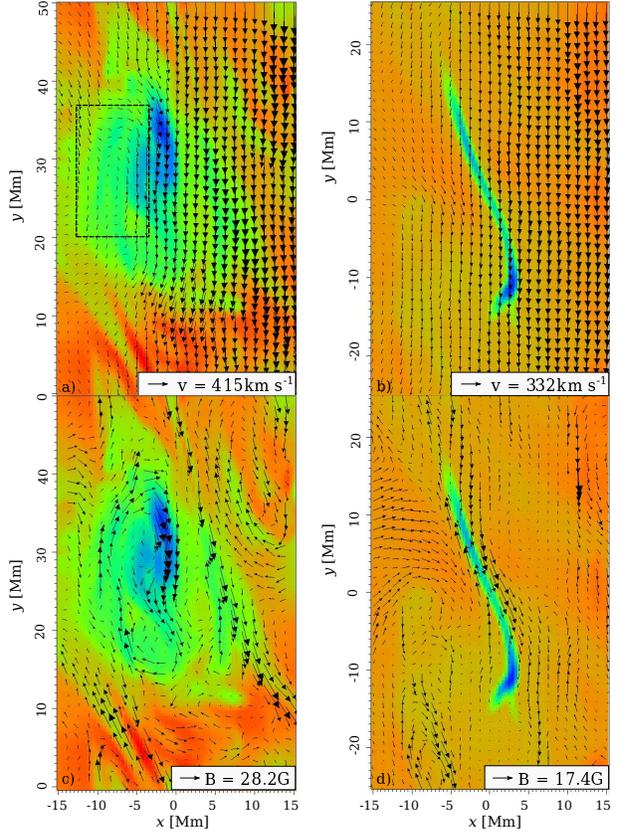}
   \caption{Velocity of the $z=0$ view for models a) $M1$ at $t=20$min and b) $M4$ at $t=49.4$min. Magnetic field  of the $z=0$ view for models c) $M1$ at $t=20$min and d) $M4$ at $t=49.4$min. The arrows inserted represent values at the scale of the vector fields for the magnitude immediately to the lower right corner.}
  \label{fig:6} 
 \end{figure}
 
 To analyze the behavior of the subdense regions,
  we show in Figure~\ref{fig:6}a--b the $z=0$ slice of the velocity
  (arrows), superimposed to the number density, for the turbulent
  vortice ($M1$) of Figure~\ref{fig:3}a and the SAD structure ($M4$)
  of Figure~\ref{fig:5}a, respectively.  Figure~\ref{fig:6}c--d is as Figure~\ref{fig:6}a--b but the arrows  are the magnetic field vector. As in
  \citet{2013ApJ...766...39M} we obtain a variable spectrum of
  velocities and vortice--like frames which correlate with the motion
  of density depletions.  
 
 The  initial  $\beta$ parameter value is $\approx 9$ and, at later times, in the fan region, ranges between $\sim[3-100]$. 
    In the vortical turbulent case (Figure~\ref{fig:6}a,c) the magnetic pressure is larger inside  than outside the vortex and the gas pressure is almost constant ($\beta_{inside}<\beta_{outside}<1$). As pointed out by \citet{2009ApJ...697.1569M, 2011ApJ...730...98S},
  Figure~\ref{fig:6}c suggests that the larger inside magnetic pressure 
 could be the reason to avoid subdense cavities to be filled in immediately by the surrounding plasma. Here,  the cavities are the subdense  eddies. 
     On the contrary, the total pressure and the   $\beta$ parameter vary smoothly around the SAD features or are almost uniform   (Figure~\ref{fig:6}b,d). 
  Figure~\ref{fig:6}d shows that the magnetic field is larger outside than inside and,  as in \citet{2012ApJ...759...79C}, the larger internal gas pressure --due to the large temperature values-- resists the filling in of the SAD. This can also be watched from movies $1$ and $2$ where the triggering and the evolution of the density and temperature of a  SAD is displayed. Note that the subdense SADs correspond to  enhanced temperature values; the pressure is almost constant.

  The eddy--like feature of
  size $\lesssim 20$Mm (indicated by a square in Figure~\ref{fig:6}a) has an average speed of $50$km s$^{-1}$. During the run we note the formation of eddies 
  of sizes ranging between $\sim[10-20]$Mm and average speeds of $[10-60]$km s$^{-1}$.
   Unlike
  the interpretation done in \citet{2012ApJ...759...79C}, where the zigzag
  behavior is due to the  interaction of 
  SADs between each other and with the inhomogeneous medium, Figure~\ref{fig:6}b,d shows that the  tail shape is 
  produced  by the interaction of a SAD with the turbulent fan.

 \subsection*{The emission measure} 

\begin{figure}[htb!]
\centering
   \includegraphics[width=6.cm]{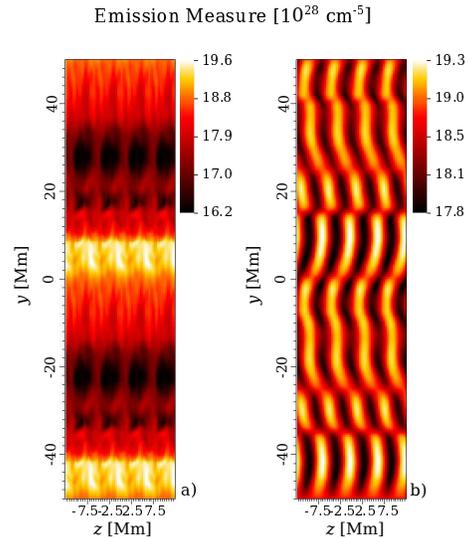}
  \caption{ Face-on view of EM for models  a) $M1$ at $20$min and b) $M2$ at $8$min.}\label{fig:7} 
 \end{figure}

 To analyze if the subdense features seen in Figures~\ref{fig:3} and
 ~\ref{fig:4} are compatible with a SAD description we first
 evaluate the emission measure (EM)
 \citep{2005psci.book.....A} as:
 \begin{equation}
EM=\int n^{2} \, dx. \label{3}
\end{equation}
  Figure~\ref{fig:7} shows the EM of
 the face-on CS view for $M1$ and $M2$. The integration along the line of sight direction ($x$) is performed considering a CS width of $4$Mm (see estimations in \citet{2013ApJ...771L..14G}). The  temperature  range obtained is $\sim[9.4-10.8]$MK for these models.
 The weak contrast obtained for the EM --where the subdense
  cavities have EMs which are less than $\sim 1.2$  the
  background features-- is lower than the reported by
  \citet{2012ApJ...747L..40S} for  $[10-13]$MK, where the EM SAD
  values  were a factor of $[2-4]$ with respect to the surroundings (see
  Figure~4 of the mentioned paper). Larger width values (e.g. considering turbulent characteristic eddy sizes as $\sim22$Mm) will lead to lower EM contrast values. Thus, the turbulent picture by itself is
  not sufficient to fully give account of dark observational regions.
 
  \begin{figure}[htb!]
\centering
   \includegraphics[width=3.cm]{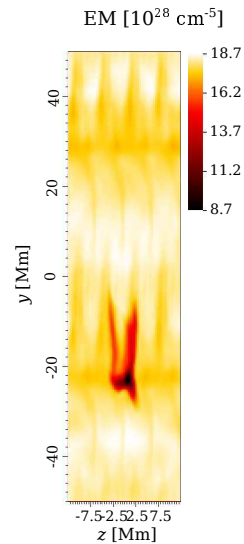}
 \caption{Face-on CS view of the EM for model $M4$, pulse size of $d=4$Mm, at  $t=50.3$min.} \label{fig:8} 
 \end{figure}
 
 Finally, Figure~\ref{fig:8} shows the CS view of the emission
 measure at  $t=50.3$min for model $M4$. 
 The EM is calculated along the line of
 sight  considering the whole temperature range ($[10-22]$MK) and assuming a thickness of the CS of $\sim 4 $Mm \citep{2013ApJ...771L..14G}. 
 At
 the initial times the SADs appear as rapidly increasing spherical
 features (e.g., see the high cadence movies of \citet{2012ApJ...747L..40S}). Later,
  as shown in Figure~\ref{fig:8}, they acquire a tailed tear--drop shape
 while they are elongated and collimated by the turbulent background
 plasma.  As stated in
 \citet{2012ApJ...747L..40S}, during times comparable with the
 observations ($\sim 270$s), the SAD emission measure is approximately $2.1$ times
 lower than its background value.  
 However, if we consider a larger CS width the SAD EM contrast  is insufficient to  be appreciated.

\begin{figure}[htb!]
\centering
   \includegraphics[width=8.cm]{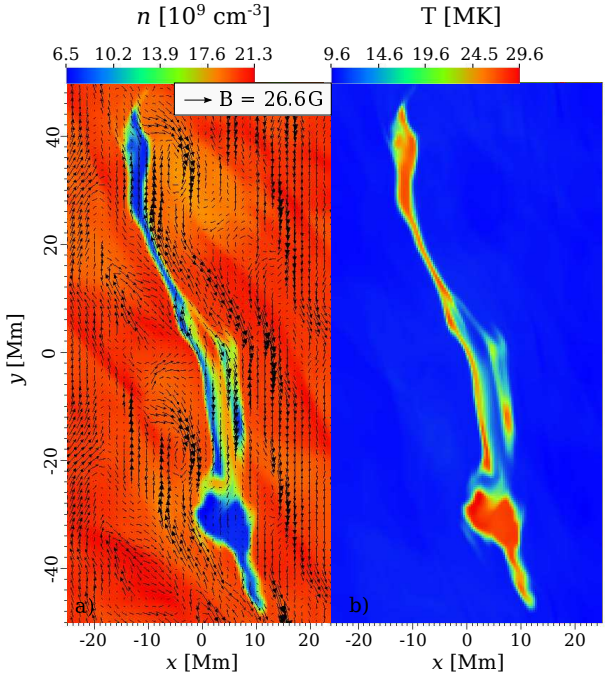}
 \caption{Slices of the $z=0$ plane for Model $M4$ with a pressure pulse diameter of $12$Mm, at $t=33$min: a) number density with the  arrows representing the magnetic field, $n$ and b) temperature, T.} \label{fig:9} 
 \end{figure}

\begin{figure}[htb!]
\centering
   \includegraphics[width=3.cm]{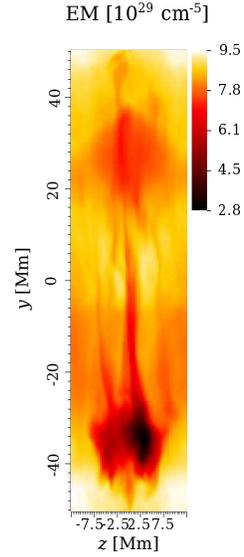}
 \caption{ Face-on CS view of the EM for model $M4$, pulse size of $d=12$Mm, at $t=33$min.} \label{fig:10} 
 \end{figure}

 Hence, we run the $M4$ model considering a CS thickness of $22$Mm to obtain a SAD using the same  instantaneous pressure pulse as before but with a  larger diameter of $12$Mm, located at $(0,30,0)$Mm and triggered at $27.5$min. This value is in accordance with a rough estimation of the eddy width in   Figure~\ref{fig:3}.
 Figure~\ref{fig:9}a-b shows the number density (with the superimposed arrows indicating the magnetic field) and the temperature, respectively for the new run, at $t=33$min. The subdense cavity (three times less than its background value, Figure~\ref{fig:9}a) sustains its structure for $5.5$min tracing a  path of $93$Mm length. The sunward SAD velocity is $\sim 280$km s$^{-1}$. The eddies  move with an average speed of  $\sim 40$km s$^{-1}$ at the neighbors of the SADs \citep{2013ApJ...766...39M}.  Figure~\ref{fig:9}b, shows larger SAD inside values of the temperature than the surrounding background. 
 In Figure~\ref{fig:10} we show the EM considering the mentioned thickness  and a temperature threshold of $[10-20]$MK taking into account the AIA temperature filters which are not sensitive to temperature higher than $20$MK  \citep{2012SoPh..275...41B}. Note that the EM contrast is sufficient ($3.4$) to be observed.  

 \section{Conclusions}\label{c}

 In this paper, motivated by recent CS observations  \citep{2013ApJ...766...39M, 2012ApJ...747L..40S},  we   simulate a turbulent CS generated by a combination of the tearing and the Kelvin--Helmholtz instabilities, that develop subdense cavities with transient and variable vortical motions. While comparing these features  with  SAD observations we find that the EM contrast and the characteristic times are not enough to match  the observations.   However,  imposing a pressure pulse to this turbulent background --in order to emulate a local deposition of energy produced by an impulsive reconnection event--, we obtain that --depending on the CS thickness--, the EM, the characteristic times and the speeds  are comparable with  the observations. 
  If the CS is thin  enough ($\sim 4$Mm), the EM contrast  will be sufficient to allow the detection of SADs. Thicker    CSs ($\sim 22$Mm) require larger depositions of energy to produce a detectable SAD. In both cases we use a triggering  pressure pulse  that  is four times the background pressure, which is a reasonable value for a pressure perturbation in a flaring medium. In the first case,   a  $d=4 $Mm size is used to obtain a detectable EM, in the second one it was required an augmented flaring region of diameter   $d=12$Mm.  These diameter values are typical  SAD sizes. 
 
 Summarizing: for appropriated physical parameters,  and a  given  pressure pulse intensity, it seems that there is a closed relation between  characteristic SAD sizes and  CS widths that must be satisfied to obtain an observable SAD. This could be a reason why SADs are not always detected during long duration flaring events.

   In contrast with the results given by  \citet{2014ApJ...786...95H}, our simulated SAD temperatures  are always higher than the fan ones. These authors pointed out that there is little evidence that SADs contain substantially hotter plasma than the surrounding fan. 
 Despite  the actual  SAD  temperatures are significantly lower than in  \citet{2011A&A...527L...5M} and \citet{2012ApJ...759...79C}, we would like  to emphasize that the scenario presented here (the scheme in Figure~\ref{fig:1}) may allow an explanation where the SAD temperature is not necessarily larger than the fan one. A more complex   
setup   simulation where a SAD is triggered  outside the fan region  (Figure~\ref{fig:1}), may lead to a SAD with  internal temperatures always higher than the surrounding background but not necessarily higher than the fan ones, e.g, if the background temperature is $T=2$MK, and the fan temperature is  $T=10$MK, the SAD initial temperature will be $T=8$MK, considering a pressure pulse of $\Delta P/P=4$. In this case the SAD would  enter the fan  while it narrows and collapse due to the total pressure difference with the fan environment, as can be also seen in movie $1b$ by  \citet{2012ApJ...747L..40S}.

 \section*{Acknowledgment}
 \noindent
 We are thankful to an anonymous referee for his/her careful reading of the paper that helped us to improve it. 

\bibliography{96041R2}
\end{document}